\documentclass[journal, twocolumn]{IEEEtran}
\usepackage{cite}
\usepackage{graphicx}
\usepackage[cmex10]{amsmath}
\usepackage{color,soul}
\usepackage{setspace}
\usepackage{amsthm,amssymb}
\usepackage{mathrsfs} 
\usepackage{algorithm,algorithmic}
\newcounter{mytempeqncnt}
\begin{document}
\title{Time-Gated Photon Counting Receivers for Optical Wireless Communication}
\author{Shenjie Huang and Majid Safari
	\thanks{The authors are with the School of Engineering, the University of Edinburgh, Edinburgh EH9 3JL, U.K. (e-mail: \{shenjie.huang, Majid.safari\}@ed.ac.uk).}
}
\maketitle
\begin{abstract}	
Photon counting detectors such as single-photon avalanche diode (SPAD) arrays are commonly considered for reliable optical wireless communication at power limited regimes. However, SPAD-based receivers suffer from significant dead-time induced intersymbol interference (ISI) especially when the incident photon rate is relatively high and the dead time is comparable or even larger than the symbol duration, i.e., sub-dead-time regime. In this work, we propose a novel time-gated SPAD receiver to mitigate such ISI effects and improve the communication performance. When operated in the gated mode, the SPAD can be activated and deactivated in well-defined time intervals. We investigate the statistics of the detected photon count for the proposed time-gated SPAD receiver. It is demonstrated that the gate-ON time interval can be optimized to achieve the best bit error rate (BER) performance. Our extensive performance analysis illustrates the superiority of the time-gated SPAD receiver over the traditional free-running receiver in terms of the BER performance and the tolerance to background light.
\end{abstract}
\section{Introduction}
In recent decades, there  has been a surge of interest in employing the photon-counting array receivers in optical wireless communications (OWC) to improve the receiver sensitivity  \cite{Chitnis:14,Fisher,Khalighi17,Huang202}. To realize a photon counting receiver, the commonly used photodiode can be biased above the breakdown voltage to operate in the Geiger mode as a single photon avalanche diode (SPAD). The SPAD-based receiver is superior to the traditional PIN and avalanche photodiode (APD) based receivers due to its single photon sensitivity and picosecond temporal resolution. It is shown that SPAD receivers can achieve sensitivity gaps to the quantum limit down to $12.7$ dB; whereas, for APD receivers the gaps are usually more than $20$ dB \cite{Zimmermann}. The application of SPAD receivers in the visible light communication (VLC) \cite{Zhang:18,Chitnis:14} and underwater wireless optical communications (UWOC) \cite{Khalighi17,Khalighi192} have been widely investigated in the literature.  However, it is well-known that the performance of the SPAD-based OWC systems is strongly limited by some non-ideal effects of SPAD, e.g., dead time, afterpulsing and crosstalk. Among these effects, dead time is probably the main limiting factor. Dead time refers to the short time period of several nanoseconds when the SPAD is unable to detect photons, which typically occurs following the avalanche introduced by each photon detection when the SPAD is getting quenched  \cite{Khalighi192}. Based on different quenching circuits, SPAD is usually classified into two main types, i.e., active quenching (AQ) and passive quenching (PQ) SPAD. For AQ SPAD the dead time remains constant; whereas, for PQ SPADs the photon arrivals during the dead time extend its duration \cite{Cova:96}. 

Due to the effects of the dead time, the average detected photon count in a symbol duration is non-linearly distorted especially in high incident photon rate regimes \cite{huang2021spad}. In addition, for high speed data transmission with symbol duration comparable or even less than the SPAD dead time, the dead time started in a symbol can extend to the subsequent symbols introducing photon counting blocking in the following symbols. Since such intersymbol interference (ISI) is inherently nonlinear, the traditional equalization techniques designed for linear channels cannot achieve optimal performance. In \cite{Huang20}, a novel detection scheme is proposed in which the information extracted from both the counts and arrival times of the detected photons are utilised for the optimal symbol detection to effectively mitigate the degradation induced by such ISI. Although SPAD receivers that can provide exact photon arrival time information are practically available \cite{patanwalareconfigurable}, having such functionality strongly increases the complexity and cost of the OWC receiver.
  
Many efforts have been devoted to the time-gated SPAD receivers mainly for imaging and spectroscopy applications to avoid the detection of unwanted photons \cite{Buller09,Tosi:11,Mora10,Cova:96}. By rising and lowering the bias voltage of the photodiode using the gate signal, the time-gated SPAD receiver can be realized in which the detector can be turned ON and OFF in well-defined time intervals \cite{Gallivanoni10}. The SPAD is working in Geiger mode only in gate-ON states and no avalanche can be triggered in gate-OFF states. In \cite{Tosi:11},  a fast-gated SPAD receiver with sub-nanosecond (less than $200$ ps) transition time from OFF to ON states and adjustable gate width down to less than $1$ ns is demonstrated to improve the dynamic range limitation of time-correlated single-photon counting (TCSPC). In addition, gating techniques are also commonly used in InGaAs SPADs to mitigate the strong afterpulsing effects \cite{Tosi12,Tosi15}. Although time-gated SPAD receivers have been applied to quantum key distribution (QKD) communications \cite{zhang2015advances}, to the best of authors' knowledge, the application of them to the OWC systems has not been investigated. In this work, we are going to employ the time-gated SPAD receiver in OWC systems to combat the  dead-time induced ISI. Through our extensive numerical results, it is demonstrated that by introducing an optimal gate-ON time window in symbol duration, the effects of ISI can be effectively mitigated and the performance of SPAD-based OWC systems can be significantly improved especially under high incident photon rates. 

The rest of this paper is organized as follows. The concept of utilizing time-gated SPAD receiver to mitigate the ISI effects is shown Section \ref{concept}. Section \ref{moments} presents the derived statistics of the received signal of time-gated SPAD receiver and its BER performance.  The numerical results and discussion are presented in Section \ref{Numer}. Finally, we conclude this paper in Section \ref{Con}.

\section{OWC with Time-Gated SPAD Receiver}\label{concept}
{Despite their large sensitivity to individual photon arrivals, due to the dead time effect, SPADs would remain blind to the incident photons for a short period after each photon detection. When SPAD is employed as OWC receiver, it might be inactive at the beginning of the symbol duration until the end of the dead time generated by the last photon arrived in the previous symbols causing a significant ISI effect \cite{Huang20}. Employing time-gated SPAD receiver can potentially mitigate such ISI effect and also reduce the unwanted detected background photon count. However, the drawback is the possible reduction of detected signal photon count. }

\begin{figure}[!t]
	\centering\includegraphics[width=0.46\textwidth]{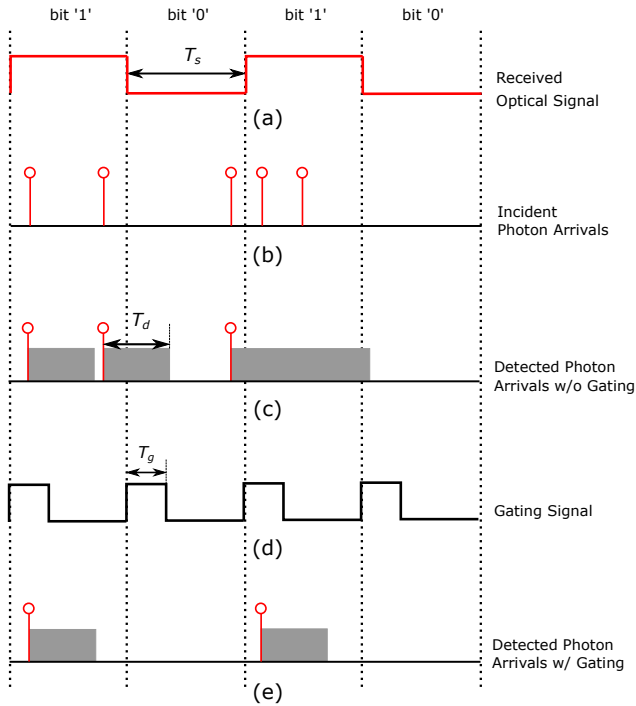}
	\caption{An example of the incident and detected photon arrivals in the presence and absence of SPAD gating functionality.} 
	\label{schematic}
%	\vspace{-5mm}
\end{figure}

To see the above points, an example of the incident and detected photon arrivals in the presence and absence of SPAD gating functionality is demonstrated in Fig. \ref{schematic}.  Figure \ref{schematic}(a) presents the received optical waveform.  In this work we consider that the transmitted signal is with OOK modulation so that optical signal received by SPAD receiver has binary photon rates. The proposed idea of using time-gated SPAD receivers can also be extended to the systems with higher order modulation schemes.  
Figure \ref{schematic}(b) shows a realization of the incident photon arrivals and the detected photon arrivals for a free-running SPAD receiver without gating is presented in Fig. \ref{schematic}(c). In this work, we assume that the employed SPAD is PQ-based and hence the photon incident during the dead time can extend the dead time duration. Note that compared to AQ SPAD, the advantages of PQ SPAD are its simpler circuit design and higher photon detection efficiency (PDE). As a result, PQ SPAD is widely employed in the commercial SPAD receivers \cite{ON}. 
It is presented in Fig. \ref{schematic}(c) that the photon arrival during the second symbol duration when a bit `0' is sent leads to an extended dead time which results in the photon arrivals in the third symbol when a bit `1' is sent undetectable. Such dead-time-induced ISI effect inevitably increases the bit error rate.

To mitigate the ISI effect caused by dead time, a gating signal comprising a sequence of short pulses with repetition rate $1/T_s$ (as shown in Fig. \ref{schematic}(d)) can be applied to the SPAD receiver where $T_s$ denotes the symbol duration. For each $T_s$, the SPAD can only detect photons during the gate duration (gate-ON time) $T_g$ and is blind to the incident photons during the rest of time period (gate-OFF time) $T_s-T_g$. In practical implementation, the gate-ON and gate-OFF time can be controlled by changing the bias voltage of the SPAD \cite{Tosi12}. For the sake of simplicity, we assume that the transition time between gate states is negligible. By introducing the gating signal, the SPAD cannot detect any photons at the end of each symbol duration. For instance, different from the free-running SPAD, the second and third incident photon arrivals shown in Fig \ref{schematic}(b) cannot be detected by time-gated SPAD as they are in the gate-OFF periods. As a result, employing time gating can effectively reduce the probability of the SPAD being inactive at the beginning of the symbol durations, which alleviates the ISI effect. For example, as presented in Fig. \ref{schematic}(e), by employing time-gated SPAD, one photon arrival in the third symbol when a bit `1' is sent can now be successfully detected, which improves the BER performance.  It is worth noting that even when the gating signal is applied, the avalanche triggered during one gate-ON time interval might still introduce a dead time that extends to the following gate-ON periods if the dead time is relatively long, which can lead to residual ISI effects.

Besides the ISI mitigation, employing time-gated SPAD can also effectively reduce the number of background photon counts and hence is beneficial to the communication performance. For instance, as shown in  Fig. \ref{schematic}, when the second symbol (bit `0') is transmitted, we expect that the received photon count is as low as possible. However, due to the existence of background light, one photon is detected during this symbol duration if SPAD receiver without gating is employed. In the presence of gating operation, since this background photon arrives during the gate-OFF time interval, it cannot be detected. 

	\begin{figure*}[!t]
	\normalsize
	\setcounter{mytempeqncnt}{\value{equation}}
	\setcounter{equation}{1}
	\begin{align}\label{uTg111}
		u(T_g)=&\int_{0}^{T_d}P\left[\mathrm{no\, arrival\, during\,}(s-T_d,0)|s \right] \!f(s)\, \mathrm{d}s+\int_{T_d}^{2T_d}P\left[\mathrm{no\, arrival\, during\,}(s-T_d,T_d)|s \right] f(s-T_d)\, \mathrm{d}s+\dots\nonumber\\[5pt]
		&+\int_{(\lfloor \frac{T_g}{T_d}\rfloor-1) T_d}^{\lfloor \frac{T_g}{T_d}\rfloor T_d}P\left[\mathrm{no\, arrival\, during\,}\left(s-T_d,(\lfloor \frac{T_g}{T_d}\rfloor-1) T_d\right)\bigg|s \right] f\left(s-(\lfloor \frac{T_g}{T_d}\rfloor-1) T_d\right)\, \mathrm{d}s\nonumber\\[5pt]
		&+\int_{\lfloor\frac{T_g}{T_d}\rfloor T_d}^{T_g}P\left[\mathrm{no\, arrival\, during\,}\left(s-T_d,\lfloor \frac{T_g}{T_d}\rfloor T_d\right)\bigg|s \right] f\left(s-\lfloor \frac{T_g}{T_d}\rfloor T_d\right)\, \mathrm{d}s.
	\end{align}
	\setcounter{equation}{\value{mytempeqncnt}}
	\hrulefill
	%	\vspace{10pt}	
\end{figure*}
\setcounter{equation}{0}
Except the aforementioned advantages of the time-gated SPAD receiver, it also has one disadvantage. Because during the gate-OFF time the signal photons are also undetectable, introducing gating functionality might result in less detected signal photon counts, which in turn degrades the performance. As presented in Fig. \ref{schematic}(c), in the absence of time gating, two signal photons can be detected in the first symbol (bit `1'); whereas, as shown in Fig. \ref{schematic}(e), only one signal photon can be detected in the presence of gating. Due to the trade-off of employing time-gated SPAD discussed above, for any given system an optimal gate ON-time $T_g^*$ should exist which can results in the best performance. In this work, this optimal gate-ON time will be investigated. 

\section{Performance of Time-Gated SPAD Receiver}\label{moments}
Different from the traditional PIN and APD photodetectors, the SPAD detectors suffer from the dead time induced non-linear distortion. 
When incident light with fixed photon rate $\lambda$ is received by a free-running PQ-based SPAD, according to the renewal theory, it is well-known that the SPAD photon transfer function is given by \cite{eisele2011185}
\begin{equation}\label{eq1}
\lambda_{D}=\lambda\,\mathrm{exp}\left(-\lambda T_d\right),
\end{equation}
where $T_d$ is the dead time and $\lambda_D$ refers to the detected photon rate. From this equation one can observe that due to the paralysis property of the PQ SPAD, with the increase of received photon rate the detected photon rate firstly increases and then decreases. The received photon rate which gives the highest detected photon rate is $1/T_d$ and the corresponding detected photon rate is $1/eT_d$. 

When SPAD is applied in OWC systems, the statistics of the detected photon count during the counting duration $T_s$ are crucial. Based on (\ref{eq1}), the average detected photon count of a free-running PQ SPAD can be expressed as $\lambda T_s\mathrm{exp}\left(-\lambda T_d\right)$ and the corresponding variance has been reported in \cite{omote,daniel2000mean}. However, when time-gated SPAD receiver is employed, the statistics of the detected photon count are not the same. In the following discussion, we will derive the mean and variance of the detected photon count for such receiver based on which the communication performance can be investigated. 

\subsection{The Statistics of the Detected Photon Count}\label{statistics}
When operated in gated mode, the SPAD is only active during the gate-ON time periods. This is equivalent to illuminate a free-running SPAD with an optical pulse wave with repetition rate $1/T_s$ and pulse width equal to the gate-ON time interval $T_g$. During gate-ON and gate-OFF intervals, the effective incident photon rates are $\lambda$ and $0$, respectively. The average detected photon count is given by the following proposition.

\newtheorem{Proposition}{Proposition}
\begin{Proposition}
Denoting the gate-ON time interval as $T_g$ with $T_g\leq T_s$, the average number of the detected photon count during $T_s$ is given by 
\stepcounter{equation}
\begin{equation}\label{ueach}
u(T_g)=\int_{0}^{\mathrm{min}\left(T_g,T_d\right)}\!\!\!\lambda \,e^{-\lambda \mathcal{G}(s)-\lambda s}\,\mathrm{d}s +\left(T_g-T_d\right)^+\lambda e^{-\lambda T_d},
\end{equation}
where
\begin{equation}\label{Gs}
\mathcal{G}(s)\!=\!\lfloor\frac{T_d-s}{T_s}\rfloor T_g+\left(T_d\!-\!s\!-\!\lfloor \frac{T_d-s}{T_s}\rfloor T_s\!-\!T_s+T_g\right)^+\!\!. 
\end{equation}
Note that $\lfloor\cdot\rfloor$ refers to the floor function and $\left(x\right)^+=\mathrm{max}\left\{x,0\right\}$. 
\end{Proposition} 
\begin{proof}
	To prove this proposition, we can consider two cases, i.e., $T_g<T_d$ and $T_g\geq T_d$. 
	
	Assuming that the investigated symbol interval starts at $t=0$ and ends at $t=T_s$, when $T_g<T_d$ holds which means at most one photon can be detected during $T_s$, the average detected photon count $u(T_g)$ can be expressed as 
	\begin{align}\label{UTg1}
	u(T_g)|&_{T_g<T_d}=\\
	&\int_{0}^{T_g}P\left[\mathrm{no\, arrival\, during\,}\left(s-T_d,0\right)\bigg|s \right] f(s)\, \mathrm{d}s.\nonumber
	\end{align}
	where $s$ denotes the time of the first photon arrival after time $t$ which according to the properties of Poisson process obeys the exponential distribution, i.e.,
	\begin{equation}\label{expo}
	f(s-t)=\lambda e^{-\lambda (s-t)}.
	\end{equation}
	Equation (\ref{UTg1}) indicates that to ensure the photon arrival at time $s$ can be successfully detected, there should be no arrival between $s-T_d$ and $s$. Note that only the first photon arrival during $[0,T_g]$ is possible to be detected since the intervals between photon arrivals within $[0,T_g]$ are always less than $T_d$, hence none of the following arrivals can be detected.  The conditional probability of no photon arrival during $(s-T_d,0)$ depends on the total gate-ON interval during this time period which is given by (\ref{Gs}). As a result, (\ref{UTg1}) can be expressed as
	\begin{equation}\label{uTgless}
	u(T_g)|_{T_g<T_d}=\int_{0}^{T_g}e^{-\lambda\mathcal{G}(s)}\lambda e^{-\lambda s} \mathrm{d}s. 
	\end{equation}
	
	\begin{figure*}[!t]
		\normalsize
		\setcounter{mytempeqncnt}{\value{equation}}
		\setcounter{equation}{11}
		\begin{align}\label{secMonfinal}
			E[\mathcal{K}^2(T_g)]=\begin{cases}
				u(T_g),& \text{if}\quad  T_g<T_d,\\[10pt]
				u(T_g)+2\,\lambda^2 e^{-\lambda T_d}\int_{0}^{T_g-T_d}e^{-\lambda\left[\mathcal{G}(s_1)+s_1\right]}(T_g-T_d-s_1) \,\mathrm{d}s_1,              & \text{if}\quad T_d\leq T_g <2T_d,\\[10pt]
				u(T_g)+\lambda^2e^{-2\lambda T_d}\left(T_g-2T_d\right)^2+2\lambda^2 e^{-\lambda T_d}\int_{0}^{T_d}e^{-\lambda\left[\mathcal{G}(s_1)+s_1\right]}(T_g-T_d-s_1)\mathrm{d}s_1, & \text{if}\quad T_g \geq 2T_d.
			\end{cases}
		\end{align}
		\setcounter{equation}{\value{mytempeqncnt}}
		\hrulefill
		%	\vspace*{4pt}
	\end{figure*}
	On the other hand, when $T_g\geq T_d$ holds, we can divide the gate opening time $T_g$ in the symbol duration $[0,T_s]$ into $\lfloor T_g/T_d\rfloor$ segments of length $T_d$ each and a remaining shorter segment with length $T_g-\lfloor T_g/T_d\rfloor T_d$. Since each segment is with length less or equal to $T_d$, at most one photon can be detected during each segment. The average photon count during $T_g$ is given by the summation of the average photon counts of these segments as given by (\ref{uTg111}). The first segment is different from the other following segments as the photon rate from $T_d$ time before until the beginning of the segment is not fixed at $\lambda$. Therefore, this segment has to be treated specially.  Similar to the case when $T_g<T_d$, the conditional probability of the first integral in (\ref{uTg111}) can be expressed as 
	\begin{equation}\label{firstint}
	P\left[\mathrm{no\, arrival\, during\,}(s-T_d,0)|s \right] =e^{-\lambda\mathcal{G}(s)},
	\end{equation}
	where $\mathcal{G}(s)$ is given by (\ref{Gs}).
	For all the following segments the probability of no arrival during $(s-T_d,t)$ can be expressed as  
	\begin{equation}\label{Pnoarr}
	P\left[\mathrm{no\, arrival\, during\,}(s-T_d,t)|s \right]=e^{-\lambda (t-s+T_d)}.
	\end{equation}
	By substituting (\ref{firstint}) and (\ref{Pnoarr}) into (\ref{uTg111}), the average photon count can be written as
	\begin{align}\label{uTg2}
	&u(T_g)|_{T_g\geq T_d}=\\[5pt]
	&\int_{0}^{T_d}\!\!e^{-\lambda\mathcal{G}(s)}\lambda e^{-\lambda s} \mathrm{d}s\!+\!\!\!\int_{T_d}^{2T_d}e^{-\lambda (2T_d-s)}\lambda e^{-\lambda (s-T_d)} \mathrm{d} s+\cdots\nonumber\\
	&+\int_{(\lfloor T_g/T_d\rfloor-1) T_d}^{\lfloor T_g/T_d\rfloor T_d}\!\!\!e^{-\lambda (\lfloor T_g/T_d\rfloor T_d-s)}\lambda e^{-\lambda \left(s-(\lfloor T_g/T_d\rfloor-1) T_d\right)} \mathrm{d} s\nonumber\\
	&+\int_{\lfloor T_g/T_d\rfloor T_d}^{T_g}e^{-\lambda (\lfloor T_g/T_d\rfloor T_d+T_d-s)}\lambda e^{-\lambda \left(s-\lfloor T_g/T_d\rfloor T_d\right)} \mathrm{d} s\nonumber.
	\end{align} 
It can be calculated that except the first term, the summation of all the other terms in the right side of (\ref{uTg2}) equals to $\left(T_g-T_d\right)\lambda e^{-\lambda T_d}$. As a result, (\ref{uTg2}) can be rewritten as
\begin{equation}\label{uTgmore}
u(T_g)|_{T_g\geq T_d}=\int_{0}^{T_d}e^{-\lambda\mathcal{G}(s)}\lambda e^{-\lambda s} \mathrm{d}s+\left(T_g-T_d\right)\lambda e^{-\lambda T_d}.
\end{equation}
Until now the average of the photon count for both $T_g<T_d$ and $T_g\geq T_d$ are derived as given in (\ref{uTgless}) and (\ref{uTgmore}), respectively. Therefore, a general expression of the average detected photon count can be expressed as (\ref{ueach}).
\end{proof}

A special case of the time-gated SPAD receiver is when $T_g=T_s$ which corresponds to the traditional free-running SPAD receiver without the gating functionality. In this case, one has $\mathcal{G}(s)=T_d-s$ and hence (\ref{ueach}) can be simplified to $u(T_g)=\lambda T_g e^{-\lambda T_d}$. The result is in line with the expression of the first moment of the detected photon count for free-running SPAD derived in  \cite{daniel2000mean,omote}.

The second moment of the detected photon count for time-gated SPAD receiver is given by the following proposition.
\begin{Proposition}\label{SecMom}
	Denoting the detected photon count of time-gated SPAD during $T_s$ as $\mathcal{K}(T_g)$, its second moment is given by (\ref{secMonfinal}).
\end{Proposition}

\begin{figure}[!t]
	\centering\includegraphics[width=0.49\textwidth]{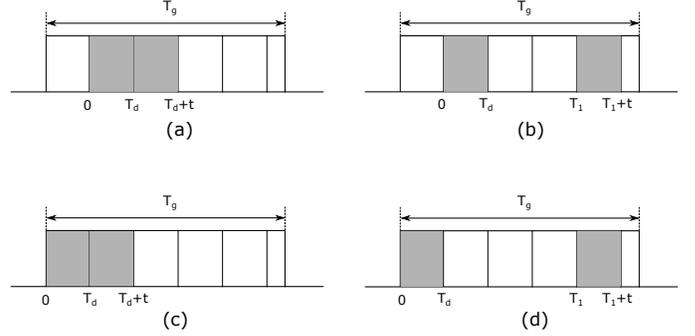}
	\caption{The four categories of the correlation terms in (\ref{secmom}): (a) and (b) refer to the cases when the two segments are adjacent and non-adjacent, respectively, and  none of the segments are the first segment; (c) and (d) refer to the corresponding cases when the first segment is involved. } 
	\label{variance}
\end{figure}	
\begin{proof}
To prove the above proposition, we can still consider two cases, i.e., $T_g<T_d$ and $T_g\geq T_d$. 	When $T_g< T_d$ holds, at most one photon can be detected during $T_g$, i.e., $\mathcal{K}(T_g)\in\{0, 1\}$. The second moment of the total detected photon count can be expressed as
\stepcounter{equation}
\begin{equation}
		E[\mathcal{K}^2(T_g)]|_{T_g<T_d}=E[\mathcal{K}(T_g)]=u(T_g)|_{T_g<T_d},
\end{equation}
where $u(T_g)|_{T_g<T_d}$ is given in (\ref{uTgless}). Note that the first equality holds since there is either zero or one photon can be detected, the second moment hence equals to the mean value.
	
Now let's turn to the case when  $T_g\geq T_d$. We can again divide the gate opening time $T_g$ within the symbol duration $[0,T_s]$ into $\lfloor T_g/T_d\rfloor$ segments of length $T_d$ each and a remaining shorter segment with length $T_g-\lfloor T_g/T_d\rfloor T_d$. Denoting the detected photon count vector of these $\lfloor T_g/T_d\rfloor+1$ segments as $\mathbf{K}=\left[K_1,K_2,\cdots,K_{\lfloor {T_g}/{T_d}\rfloor+1}\right]$, the second moment of the total detected photon count is given by 
\stepcounter{equation}
\begin{align}\label{secmom}
E[\mathcal{K}^2(T_g)]&=E\bigg[\bigg(\sum_{n=1}^{\lfloor {T_g}/{T_d}\rfloor+1}K_n\bigg)^2\bigg]\\
&=\!\!\sum_{n=1}^{\lfloor {T_g}/{T_d}\rfloor+1} \!\!\!E\left[K_n^2\right]\!\!+\!2\!\!\!\sum_{n=1}^{\lfloor {T_g}/{T_d}\rfloor}\sum_{j=n+1}^{\lfloor {T_g}/{T_d}\rfloor+1}\!\!\!\!E\left[K_nK_j\right].\nonumber
\end{align}
%	When $T_g< T_d$, there is only one segment within $T_g$. As a result, the second correlation term in (\ref{secmom}) is removed and (\ref{secmom}) can be rewritten as 
%	\begin{equation}
%	E[\mathcal{K}^2(T_g)]|_{T_g<T_d}=E[K_1^2]=E[K_1]=u(T_g)|_{T_g<T_d},
%	\end{equation}
%	where $u(T_g)|_{T_g<T_d}$ is given in (\ref{uTgless}). Note that the second equality holds since there is either zero or one photon can be detected, the second moment hence equals to the mean value.
Since each segment is with length less or equal to the dead time, the first term of (\ref{secmom}) equals to the average value $u(T_g)|_{T_g\geq T_d}$ given in  (\ref{uTgmore}). As mentioned when the first moment is derived, the first segment differs from the following segments, since the photon rate $T_d$ time before until the beginning of this segment is not fixed at $\lambda$. Therefore the correlation terms in (\ref{secmom}) can be classified into four categories as shown in Fig. \ref{variance}. Note that Fig. \ref{variance}(a) and Fig. \ref{variance}(b) refer to the cases when the two segments are adjacent and non-adjacent, respectively, under the condition that neither of the segments are the first segment; Fig. \ref{variance}(c) and Fig. \ref{variance}(d) refer to the corresponding cases when the first segment is involved. When calculating the correlation between two segments, the preceding segment $K_n$ is always with duration $T_d$; whereas, the latter segment $K_j$ could be with duration less than $T_d$ when it is the last segment within $T_g$. Therefore, as shown in Fig \ref{variance}, we denote the duration of the latter segment as $t$ with $t\in\{T_d,T_g-\lfloor {T_g}/{T_d}\rfloor T_d\}$. Let us define the correlation function $E\left[K_nK_j\right]$ of these four cases as $\psi_{A}(t)$, $\psi_{B}(t)$, $\psi_{C}(t)$, $\psi_{D}(t)$, respectively. Based on  the above observations, when $T_d\leq T_g <2T_d$ holds, (\ref{secmom}) contains only one correlation term which is in the category of (c). As a result, the second moment (\ref{secmom}) can be rewritten as 
	\begin{equation}\label{secMom3}
	E[\mathcal{K}^2(T_g)]|_{T_d\leq T_g <2T_d}=u(T_g)|_{T_g\geq T_d}+2\,\psi_{C}\left(T_g-T_d\right),
	\end{equation}
	where $u(T_g)|_{T_g\geq T_d}$ is given by (\ref{uTgmore}). On the other hand, when $T_g\geq 2T_d$ holds, the corresponding second moment can be expressed as   
	\begin{align}\label{secMom1}
	&\!\!\!\!\!\!\!\!\!\!\!\!E[\mathcal{K}^2(T_g)]|_{T_g\geq 2T_d}=u(T_g)|_{T_g\geq T_d}\\[5pt]
	&+2\left[\left(\lfloor \frac{T_g}{T_d}\rfloor-2\right)\psi_A(T_d)+\psi_{A}\left(T_g-\lfloor \frac{T_g}{T_d}\rfloor T_d\right)\right.\nonumber
	\end{align}
	\begin{align}
	&+\!\!\frac{\left(\!\lfloor \frac{T_g}{T_d}\rfloor\!\!-\!2\!\right)\!\!\left(\!\lfloor \frac{T_g}{T_d}\rfloor\!\!-\!3\!\right)}{2}\psi_{B}(T_d)\!+\!\!\left(\!\!\lfloor \frac{T_g}{T_d}\rfloor\!\!-2\!\right)\!\!\psi_{B}\!\left(\!\!T_g\!-\!\lfloor \frac{T_g}{T_d}\rfloor T_d\!\right)\nonumber\\	
	&\left.+\psi_{C}\left(T_d\right)+\left(\lfloor \frac{T_g}{T_d}\rfloor-2\right)\psi_{D}(T_d)+\psi_{D}(T_g-\lfloor \frac{T_g}{T_d}\rfloor T_d)\right].\nonumber
	\end{align}
	In the following discussion, the correlation functions $\psi_{A}(t)$, $\psi_{B}(t)$, $\psi_{C}(t)$, and $\psi_{D}(t)$ will be derived. 
	
	For case (a), denoting the photon arrival in the two segments as $s_1$ and $s_2$, respectively, the correlation function is given by
	\begin{align}\label{casea}
	\psi&_{A}(t)\!=\!\!\int_{0}^{t}\!\!\int_{s_1+T_d}^{T_d+t}\!\!P\left[\mathrm{no\, arrival\, during\,}(s_1\!-\!T_d,0)|s_1 \right] \!f(s_1)\nonumber\\[5pt]
	&\cdot P\left[\mathrm{no\, arrival\, during\,}(s_2\!-\!T_d,T_d)|s_2 \right] \!f(s_2\!-\!T_d)\,\mathrm{d}s_2\,\mathrm{d}s_1. 
	\end{align}      
	Considering that the probability of no photon arrival for a time period of $t$ is given by $e^{-\lambda t}$ and the definition of function $f(s-t)$ given in (\ref{expo}), equation (\ref{casea}) can be rewritten as 
	\begin{align}\label{psiA}
	&\psi_{A}(t)\\[5pt]
	&=\int_{0}^{t}\!\!\int_{s_1+T_d}^{T_d+t}\!\!\!e^{-\lambda(T_d-s)}\lambda  e^{-\lambda s_1}e^{-\lambda(2T_d-s_2)} \lambda e^{-\lambda(s_2-T_d)}\mathrm{d}s_2\mathrm{d}s_1\nonumber\\[5pt]
	&=\frac{1}{2} \lambda^2t^2e^{-2\lambda T_d}\nonumber.
	\end{align}
	For case (b), the correlation function is given by 
	\begin{align}\label{caseb}
	&\psi_{B}(t)=\\
	&\int_{0}^{T_d}\int_{T_1}^{T_1+t}P\left[\mathrm{no\, arrival\, during\,}(s_1-T_d,0)|s_1 \right] \!f(s_1)\nonumber\\
	&\cdot P\left[\mathrm{no\, arrival\, during\,}(s_2-T_d,T_1)|s_2\right] \!f(s_2-T_1)\,\mathrm{d}s_2\,\mathrm{d}s_1. \nonumber
	\end{align} 
	Note that since the considered two segments are non-adjacent and hence with interval larger than $T_d$, the detected photon counts in these two segments are independent. Therefore, (\ref{caseb}) can be simplified as the product of two integrals
	\begin{align}\label{caseb2}
	&\psi_{B}(t)\\
	&= \!\!\!\int_{0}^{T_d}\!\!\!\!\!e^{-\lambda(T_d-s_1)}\lambda e^{-\lambda s_1}\mathrm{d}s_1\!\!\!\int_{T_1}^{T_1+t}\!\!\!\!\!\!\!\!e^{-\lambda(T_1-s_2+T_d)}\lambda e^{-\lambda (s_2-T_1)}\mathrm{d}s_2\nonumber\\
	&=\lambda^2T_d\, t\,e^{-2\lambda T_d}.\nonumber
	\end{align} 
	For case (c), the correlation function can still be presented as (\ref{casea}); however, since the photon rate before the beginning of the first segment is not fixed at $\lambda$,  the conditional probability of no arrival during $(s_1-T_d,0)$  does not simply equal to $e^{-\lambda(T_d-s_1)}$ but equals to $e^{-\lambda\mathcal{G}(s_1)}$ with $\mathcal{G}(s)$ given in (\ref{Gs}). Therefore, the correlation function for this case can be expressed as 
	\begin{align}\label{psiC}
	&\psi_C(t)\nonumber\\
	&=\!\!\int_{0}^{t}\!\!\!\int_{s_1+T_d}^{T_d+t}\!\!e^{-\lambda\mathcal{G}(s_1)}\lambda  e^{-\lambda s_1}e^{-\lambda(2T_d-s_2)} \lambda e^{-\lambda(s_2-T_d)}\,\mathrm{d}s_2\,\mathrm{d}s_1\nonumber\\
	&=\lambda^2 e^{-\lambda T_d}\int_{0}^{t}e^{-\lambda\left[\mathcal{G}(s_1)+s_1\right]}(t-s_1) \,\mathrm{d}s_1.
	\end{align}
	Finally, for case (d), the correlation function is given by 
	\begin{align}\label{psiD}
	&\psi_{D}(t)\nonumber\\
	&=\!\!\! \int_{0}^{T_d}\!\!\!\!e^{-\lambda\mathcal{G}(s_1)}\lambda e^{-\lambda s_1}\mathrm{d}s_1\!\!\int_{T_1}^{T_1+t}\!\!\!\!\!e^{-\lambda(T_1-s_2+T_d)}\lambda e^{-\lambda (s_2-T_1)}\mathrm{d}s_2\nonumber\\
	&=\lambda\,t\,e^{-\lambda T_d}\int_{0}^{T_d}e^{-\lambda\mathcal{G}(s_1)}\lambda e^{-\lambda s_1}\mathrm{d}s_1.
    \end{align} 
	
	By substituting (\ref{psiC}) into (\ref{secMom3}), the second moment of the detected photon count when $T_d\leq T_g <2T_d$ can be rewritten as $E[\mathcal{K}^2(T_g)]|_{T_d\leq T_g <2T_d}=$
	\begin{equation}\label{secMom32}
	u(T_g)|_{T_g\geq T_d}+2\,\lambda^2 e^{-\lambda T_d}\!\!\!\!\int_{0}^{T_g-T_d}\!\!\!\!e^{-\lambda\left[\mathcal{G}(s_1)+s_1\right]}(T_g\!-T_d-s_1) \,\mathrm{d}s_1.
	\end{equation} 
	On the other hand, by substituting (\ref{psiA}), (\ref{caseb2}), (\ref{psiC}) and (\ref{psiD}) into (\ref{secMom1}), the second moment of the detected photon count when  $T_g \geq2T_d$ can be expressed as
	\begin{align}\label{secMom4}
	E[\mathcal{K}^2(T&_g)]|_{T_g\geq 2T_d}=\\
	&u(T_g)|_{T_g\geq T_d}+\lambda^2e^{-2\lambda T_d}\left(T_g-2T_d\right)^2\nonumber\\
	&+2\lambda^2 e^{-\lambda T_d}\int_{0}^{T_d}e^{-\lambda\left[\mathcal{G}(s_1)+s_1\right]}(T_g-T_d-s_1) \,\mathrm{d}s_1.\nonumber
	\end{align}
	In summary, the second moment of the detected photon count is presented by (\ref{secMonfinal}). 
\end{proof}

We can also consider a special case when the gating time $T_g=T_s$ which corresponds to the traditional free-running SPAD receiver. Since in this scenario $\mathcal{G}(s)=T_d-s$ and $u(T_g)=\lambda T_s e^{-\lambda T_d}$, by substituting these terms into (\ref{secMonfinal}) and after some mathematical manipulations, one can get the second moment of the detected photon count as $E[\mathcal{K}^2(T_g)]=$
\begin{equation}
\begin{cases}
\lambda T_g e^{-\lambda T_d},& \text{if}\quad  T_g<T_d,\\
\lambda T_g e^{-\lambda T_d}+\lambda^2 e^{-2\lambda T_d}\left(T_g-T_d\right)^2, &\text{if}\quad  T_g\geq T_d.\nonumber
\end{cases}
\end{equation}
This result is in line with the second moment expression of the detected photon count presented in  \cite{omote}. 

With the derived first and second moments of the detected photon count given in (\ref{ueach}) and (\ref{secMonfinal}) respectively, its variance as a function of $T_g$ can be expressed as 
\begin{equation}\label{varieach}
\sigma^2(T_g)=E[\mathcal{K}^2(T_g)]-u^2(T_g). 
\end{equation}
\begin{figure}[!t]
	\centering\includegraphics[width=0.5\textwidth]{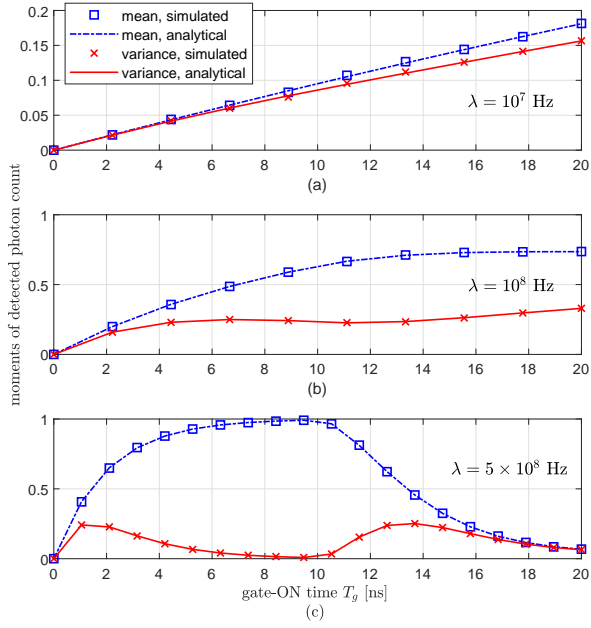}
	\caption{The mean and variance of the detected photon count versus the gate-ON time $T_g$ for time-gated SPAD under various incident photon rate $\lambda$ with dead time $T_d=10$ ns and symbol time $T_s=20$ ns. } 
	\label{moments_figure}
\end{figure}

An example of the mean and variance of the detected photon count during $T_s$ versus $T_g$ under various photon rate $\lambda$ is plotted in Fig. \ref{moments_figure}. It is demonstrated that the derived analytical results shown in (\ref{ueach}) and (\ref{varieach}) excellently match with the simulation results, which justifies our analytical derivations. When the photon rate is relatively low, e.g., $\lambda=10^7$ Hz and $\lambda=10^8$ Hz, with the increase of $T_g$, the mean detected photon count monotonically increases as presented in Fig. \ref{moments_figure}(a) and Fig. \ref{moments_figure}(b). This is because under low photon rates the dead time induced ISI is negligible and increasing $T_g$ is always beneficial which increases the probability of detecting photons. The above observation is not the case when the incident photon rate is relatively high, e.g., $\lambda=5\times10^8$ Hz. In this scenario, as shown in Fig. \ref{moments_figure}(c), with the increase of $T_g$, the average detected photon count firstly increases; but when $T_g$ goes beyond the dead time $T_d=10$ ns, the average photon count gradually reduces. This is because for a counting period of $T_s=20$ ns, when $T_g$ is less than $10$ ns, the gate-OFF time interval is larger than the dead time $T_d=10$ ns. As a result, the photon detection in one gate-ON interval will not introduce any ISI effects to the following gate-ON intervals, hence increasing $T_g$ gives higher average photon count. When $T_g$ is close to $10$ ns the mean value of detected photon count approaches $1$ and the corresponding variance approaches $0$, since the high photon rate and the absence of ISI guarantee that one photon can always be detected during $T_g$. But when $T_g$ is larger than $10$ ns, significant ISI effect comes into place which increases the probability of SPAD being blocked during the gate-ON time intervals due to the previous triggered avalanches. Higher $T_g$ could lead to more severe ISI effects and strongly reduce the detected photon count. Therefore, the average detected photon count reduces with the further increase of $T_g$.    

\subsection{The BER Performance}\label{BERper}
\begin{figure}[!t]
	\centering\includegraphics[width=0.5\textwidth]{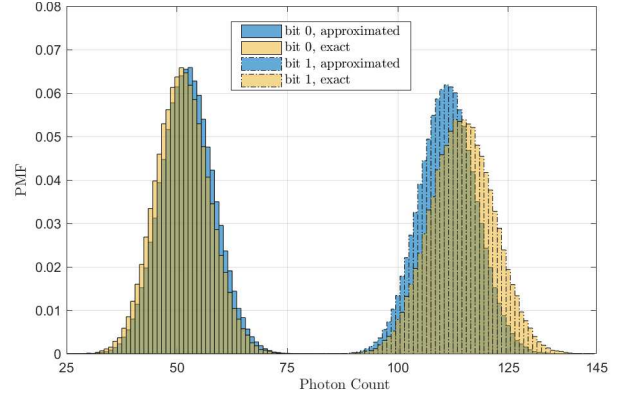}
	\caption{The exact and approximated conditional PMFs of the detected photon count when bit `0' and bit `1' are transmitted where $\lambda_0=2\times 10^7$ Hz, $\lambda_1=7\times 10^7$ Hz, $T_d=10$ ns, $T_s=50$ ns and the number of SPAD $N=64$.  } 
	\label{PDF_issue}
\end{figure}
{When modulated OOK signal is received, for a SPAD array detector with $N$ pixels the received photon rate for each pixel when bit `0' and bit `1' are sent can be expressed as  
\begin{align}\label{lam0lam1}
	\lambda_0&=\frac{\Upsilon_\mathrm{PDE} P_b}{Nh\nu},  \nonumber\\ \lambda_1&=\frac{\Upsilon_\mathrm{PDE} \left(2P_R+P_b\right)}{Nh\nu}. \bibliographystyle{IEEEtran}
\end{align}   
where $P_R$ and $P_b$ denote the received signal power and background light power, respectively, $\Upsilon_\mathrm{PDE}$ is the PDE of the SPAD, $h$ refers to the Planck constant, and $\nu$ denotes the light frequency. By substituting $\lambda_0$ and $\lambda_1$ into (\ref{ueach}) and (\ref{varieach}), the approximated first two moments of the detected photon counts when bit `0' and bit `1' are received can be calculated. 
We denote that when bit `1' is sent, the mean and variance of the detected photon count during $T_s$ are $u_1(T_g)$ and $\sigma_1^2(T_g)$, respectively; whereas when bit `0' is sent the corresponding moments are  $u_0(T_g)$ and $\sigma_0^2(T_g)$. When the size of the SPAD array is relatively large, due to the central limit theorem, for bit `1' (bit `0') the received photon count can be modelled as Gaussian distribution with mean $Nu_1(T_g)$ ($Nu_0(T_g)$) and variance $N\sigma_1^2(T_g)$ ($N\sigma_0^2(T_g)$) \cite{Khalighi17}. As a result, the BER can be expressed as \cite{elham152}
\begin{equation}\label{BER}
	BER_\mathrm{SPAD}(T_g)=Q\left[\frac{\sqrt{N}u_1(T_g)-\sqrt{N}u_0(T_g)}{\sigma_1(T_g)+\sigma_0(T_g)}\right],
\end{equation}
where $Q[\cdot]$ refers to the Q-function. The optimal $T_g$, denoted as $T_g^*$, which results in the minimum BER can be found through exhaustive search. Note that the BER of the free-running SPAD receiver can also be achieved by using (\ref{BER}) with $T_g=T_s$. }

{It is worth noting that the above calculated moments of detected photon count are approximated ones, because the moments derived in Section \ref{statistics} are calculated considering that the incident optical signal is unmodulated. The exact statistics of the detected photon count when modulated light is received are mathematically intractable. The same approximation has been widely employed in the literature. For example, inspired by the photon transfer function in (\ref{eq1}), some works in the literature assume that when bit `0' with photon rate $\lambda_0$ and bit `1' 
with photon rate $\lambda_1$ are received, the averaged detected photon counts during $T_s$ are  $\lambda_0T_se^{-\lambda_0T_d}$ and $\lambda_1T_se^{-\lambda_1T_d}$, respectively, and the BER performance is investigated based on this assumption \cite{Khalighi192,Li15,Khalighi17,Zou19}. }

{By employing the approximated moments, the calculated BER (\ref{BER}) would be a bound on the exact BER performance. Figure \ref{PDF_issue} presents the difference between the exact and approximated PMF of the detected photon count when modulated signal is received. In fact, by employing the approximation, it is assumed that when the current received bit is bit `1', all of the previously transmitted bits are also bit `1'. Since bit `1' is with photon rate higher than bit `0', employing such approximation overestimates the dead-time induced ISI and hence underestimates the photon count when bit `1' is transmitted. This can be observed in Fig. \ref{PDF_issue} by comparing the PMFs of the photon count when bit `1' is sent in the presence and absence of the approximation. In contrast, as also presented in Fig. \ref{PDF_issue}, this approximation overestimates the detected photon count for bit `0', since less dead-time induced ISI degradation is considered.    
From the error rate perspective, when the approximated moments are employed, the underestimation of the photon count for bit `1' and overestimation of the photon count for bit `0' together result in a lower bound of the real SPAD BER performance. However, as later demonstrated in the numerical results in Section  \ref{Numer}, the gap between the approximated BER (\ref{BER}) and the exact one is small when the dead time is less or comparable to the symbol duration. }

\section{Numerical Results}\label{Numer}
In this section, some numerical results regarding the time-gated SPAD receiver are presented. The optical wavelength is considered as $785$ nm, the PDE of SPAD is set as $\Upsilon_\mathrm{PDE}=0.18$, and the dead time is $T_d=10$ ns. The considered SPAD receivers are with two array sizes, i.e., $N=64$ and $N=1024$. The selection of the optimal gate-ON time is firstly discussed and the superiority of the time-gated SPAD receiver over the traditional free-running receiver in terms of the BER performance is  investigated.  
\begin{figure}[!t]
	\centering\includegraphics[width=0.48\textwidth]{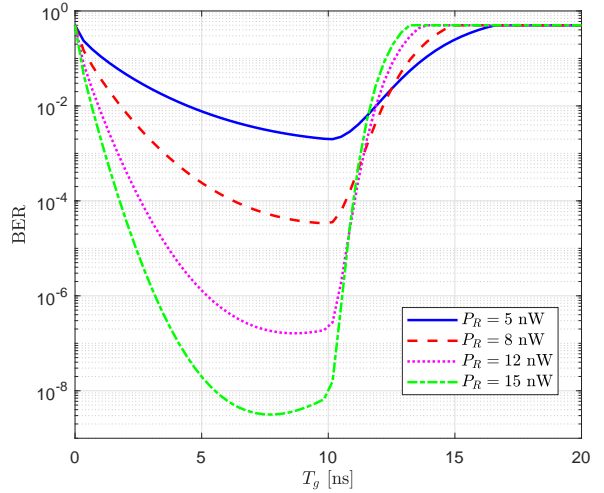}
	\caption{BER versus the gate-ON interval $T_g$ for a small time-gated SPAD array with $N=64$ under various received signal power $P_R$ when $P_b=7$ nW and data rate $R=50$ Mbps.} 
	\label{BER_Tg}
\end{figure}
	\begin{figure}[!t]
	\centering\includegraphics[width=0.5\textwidth]{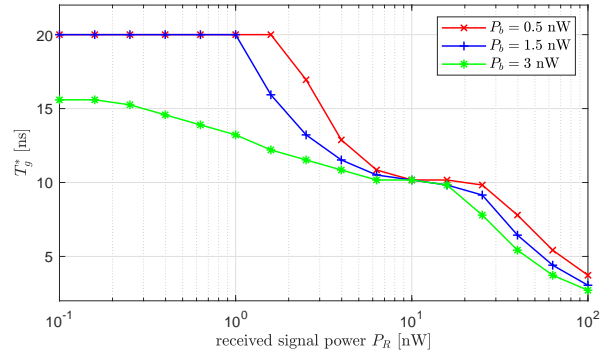}
	\caption{The optimal $T_g$ versus the received signal power $P_R$ for a small SPAD array with $N=64$ and data rate $R=50$ Mbps under various background light power $P_b$. } 
	\label{opt_Tg_64}
\end{figure}

	Fig. \ref{BER_Tg} plots the BER versus the gate-ON time interval $T_g$ for a small SPAD array with $N=64$ under various $P_R$. For such a small array size, the data rate is quite limited, hence we assume a data rate of $R=50$ Mbps which corresponds to a symbol duration of $T_s=20$ ns. Invoking the dead time $T_d=10$ ns, this symbol duration is twice of the dead time. It is demonstrated that for any given $P_R$, with the increase of $T_g$, BER firstly decreases due to higher probability of detecting signal photons and then increases mainly because of the severer ISI effects. This trade-off results in an optimal gate-ON interval $T_g^*$ that achieves the minimal BER. For instance, when $P_R=8$ nW and $P_R=15$ nW, the minimal BER is $3.5\times10^{-5}$ with $T_g^*=10$ ns and $3.1\times10^{-9}$ with $T_g^*=7.8$ ns, respectively.  Note that when $T_g$ goes above $10$ ns, the gate-OFF interval $T_s-T_g$ becomes less than $T_d$ and the dead-time induced ISI effects appear. As a result, the communication performance would be significantly degraded and the BER suddenly increases as presented in Fig. \ref{BER_Tg}. In addition, it is shown in Fig. \ref{BER_Tg} that in the absence of ISI effects, i.e., $T_g<10$ ns, larger $T_g$ increases the number of detected signal photon count but this does not necessarily indicate better performance, for example when $P_R=12$ nW and $P_R=15$ nW. This is because higher $T_g$ also increases the number of detected background photon counts which could in turn degrade the performance.
	   
	For the system considered in Fig. \ref{BER_Tg}, the optimal $T_g$ under various $P_R$ and $P_b$ is presented in Fig. \ref{opt_Tg_64}.  It is demonstrated that with the increase of $P_R$, the optimal gate-ON time interval $T_g^*$ decreases. For instance, with $P_b=0.5$ nW, the optimal $T_g$ is $20$ ns when $P_R=1$ nW which refers to a free-running SPAD receiver. However, the corresponding optimal $T_g$ drops to $10$ ns when $P_R$ increases to $10$ nW. The reason behind this is that larger $P_R$ means stronger dead time induced ISI effects which requires a lower $T_g$ to mitigate the corresponding degradation. In addition, it is also demonstrated that systems with higher background light power $P_b$ has relatively lower $T_g^*$. This is due to the fact that in the presence of relatively high background light power, the system performance is strongly limited by the detected background photon count. As a result, choosing a lower $T_g$ which effectively decreases the detected background photon counts can  improve the performance.   
	
\begin{figure}[!t]
	\centering\includegraphics[width=0.5\textwidth]{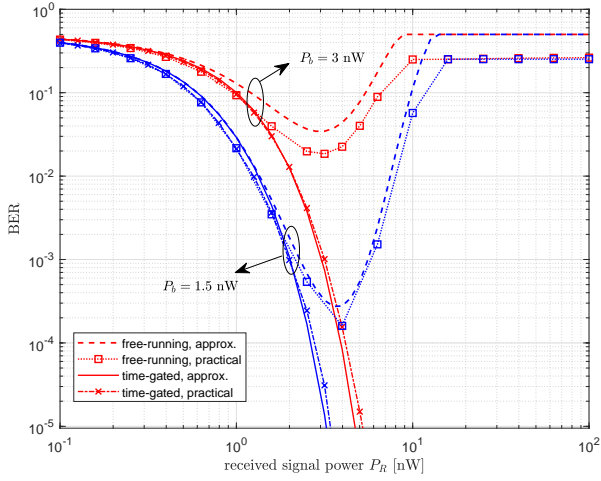}
	\caption{BER versus the received signal power $P_R$ for a small SPAD array with $N=64$ and data rate $R=50$ Mbps under various background light power $P_b$ in the presence and absence of the gating operation.} 
	\label{BER_PR_64}
\end{figure}

Figure \ref{BER_PR_64} presents the BER versus $P_R$ for $N=64$ under various $P_b$ for time-gated and free-running SPAD receivers. For time-gated receiver the optimal $T_g$ which minimizes the BER is adopted. In this figure, besides the approximated BER performance calculated based on (\ref{BER}), the exact performance of the practical SPAD receivers is also plotted. This performance is achieved
by firstly generating the random transmitted data stream based on which the long photon arrival sequence received by each SPAD pixel can be generated by using the theory of Poisson process; then filtering the photon arrival sequences using the dead time; calculating the total photon count detected by the array in every symbol duration; and finally decoding the signal based on the detected photon counts. The above process can exactly mimic the photon reception of the practical SPAD receiver, but it is with high computational complexity. It is shown in Fig. \ref{BER_PR_64} that for free-running receivers without the gating operation, with the increase of $P_R$ the BER firstly decreases and then increases due to the stronger dead-time induced ISI. In addition, the BER performance is very sensitive to the background light intensity. For instance, when $P_b$ is $1.5$ nW, a BER low than $10^{-4}$ cannot be achieved no matter what is $P_R$ and when $P_b$ increase to $3$ nW, even a BER of $10^{-2}$ cannot be achieved. It is also demonstrated in Fig. \ref{BER_PR_64} that the approximated BER performance is slightly worse than that of the practical receiver as mentioned in Section \ref{BERper}, but the performance gap is relatively small. 

It is illustrated in Fig. \ref{BER_PR_64} that the time-gated receiver with $T_g^*$ strongly outperforms the free-running one especially in high $P_R$ regimes. For instance, when $P_R=4$ nW and $P_b=3$ nW, by using free-running receiver, the achievable BER is only $0.04$, but the corresponding BER drops to $8.4\times10^{-5}$ when time-gated receiver is employed. Furthermore, Fig. \ref{BER_PR_64} illustrates that by limiting and even totally removing the dead-time induced ISI through adjusting $T_g$, time-gated SPAD can achieve a monotonic decrease of BER with the increase of $P_R$. Time-gated receiver also has higher tolerance to the background light compared to its counterpart. For example, when $P_b$ is $1.5$ nW and $3$ nW, a BER less than $10^{-4}$, which can not be achieved for free-running receiver, can always be guaranteed for time-gated receiver when $P_R$ is higher than $2.6$ nW and $4$ nW, respectively. In addition, one can observe that for time-gated receiver the approximated and practical BER results are very close. Note that the considered $T_g^*$ which is calculated based on the approximated BER (\ref{BER}) is the optimal gate-ON time under the approximation mentioned in Section \ref{BERper}, but is not necessarily the optimal gate-ON time for the practical SPAD receiver. As a result, different from the free-running receiver, the approximated performance is not always worse than the performance of the practical receiver.   
    
\begin{figure}[!t]
	\centering\includegraphics[width=0.48\textwidth]{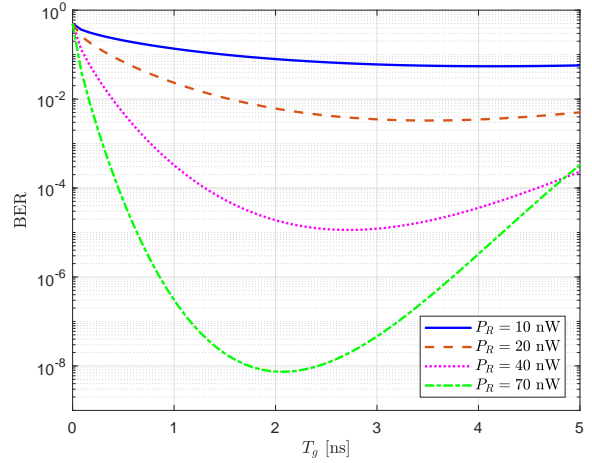}
	\caption{BER versus $T_g$ for a large time-gated SPAD array with $N=1024$ under various received signal power $P_R$ when $P_b=40$ nW and data rate $R=200$ Mbps.} 
	\label{BER_Tg2}
\end{figure}
\begin{figure}[!t]
	\centering\includegraphics[width=0.5\textwidth]{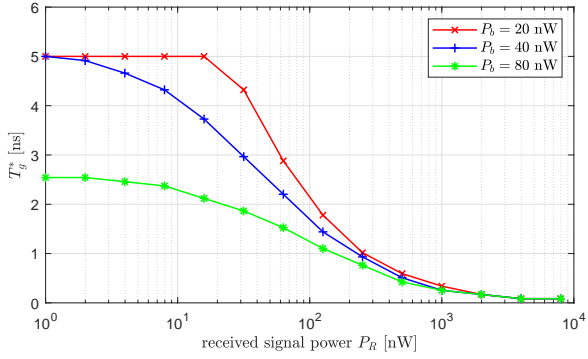}
	\caption{The optimal $T_g$ versus the received signal power $P_R$ for a large SPAD array with $N=1024$ and data rate  $R=200$ Mbps under various background light power $P_b$. } 
	\label{opt_Tg}
\end{figure}
Now let's turn to the SPAD receiver with a larger array size, i.e., $N=1024$. With larger array size the dead time effects are  mitigated and the receivers are capable of operating under higher data rate $R$, signal power $P_R$ and background light power $P_b$. We assume that the data rate is $R=200$ Mbps which corresponds to a sub-dead-time scenario with the symbol duration $T_s=5$ ns half of the dead time. Fig.  \ref{BER_Tg2} presents the BER versus $T_g$ under various $P_R$ when $P_b=40$ nW. Similar to Fig. \ref{BER_Tg}, it is again shown that for any given $P_R$, the optimal $T_g$ can always be determined. For instance, when $P_R=40$ nW and $P_R=70$ nW, the optimal gate-ON interval $T_g^*$ are $2.7$ ns and $2$ ns, respectively. However, different from the system considered in Fig. \ref{BER_Tg} where the ISI effects can be thoroughly eliminated by reducing $T_g$, here the ISI effects cannot be totally removed, since the gate-OFF interval is always shorter than $T_d$. Therefore, a sharp increase of BER with the increase of $T_g$ due to the appearance of ISI effects cannot be observed in Fig.  \ref{BER_Tg2}.   The relationship between optimal $T_g$ and $P_R$ is demonstrated in Fig. \ref{opt_Tg} which again confirms that with the increase of $P_R$, $T_g^*$ monotonically decreases and larger $P_b$ results in lower $T_g^*$. 
\begin{figure}[!t]
	\centering\includegraphics[width=0.5\textwidth]{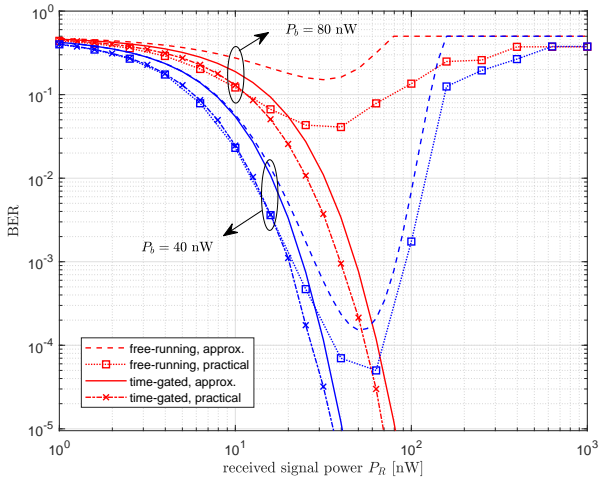}
	\caption{BER versus the received signal power $P_R$ for a large SPAD array with $N=1024$ and data rate $R=200$ Mbps under various background light power $P_b$ in the presence and absence of the gating operation.} 
	\label{BER_PR}
\end{figure}

The BER versus $P_R$ for the large SPAD array under various $P_b$ is shown in Fig. \ref{BER_PR}. Similar to Fig. \ref{BER_PR_64}, it can be observed that the time-gated SPAD receiver significantly outperforms the free-running receiver in terms of BER performance and background light tolerance. For example, with $P_b=80$ nW and $P_R=63$ nW, by employing time-gated SPAD receiver, a BER of $10^{-4}$ can be achieved; however, the corresponding BER of free-running receiver is around $0.1$. It is worth noting that the performance gaps between the approximated and the practical BER shown in Fig. \ref{BER_PR} are larger than those in Fig. \ref{BER_PR_64} due to the smaller considered symbol duration to dead time ratio.

%\begin{figure}[!t]
%	\centering\includegraphics[width=0.48\textwidth]{DR_PR_new_v2.eps}
%	\caption{The achievable data rate versus the received signal power $P_R$ for SPAD array with $N=2048$ under various background power $P_b$ in the presence and absence of the gating operation. The BER target is set as $P_{e,th}=10^{-3}$.  } 
%	\label{DR_PR_all}
%\end{figure}
%Finally, in order to provide some insights on the improvement of achievable data rate by employing time-gated SPAD receiver, Fig. \ref{DR_PR_all} is plotted where the array size is $N=2048$ and the BER  target is $P_{e,th}=10^{-3}$. For free-running receiver, due to the dead-time induced ISI effects, with
%the increase of PR the achievable data rate firstly increases and then decreases. The maximal achievable data rate is $2$ Gbps when $P_b=30$ nW and it reduces to $300$ Mbps when $P_b=100$ nW. On the other hand, when the SPAD operated in time-gated mode, with the increase of $P_R$, the achievable data rate keeps increasing and becomes dramatically higher than that of free-running receiver. For instance, with $P_R=400$ nW and $P_b=30$ nW, the achievable data rate of free-running receiver is only $204$ Mbps; whereas, by using time-gated receiver, the achievable data rate significantly increases to $4$ Gbps.    
 
\section{Conclusion}\label{Con}
In this work, we propose to employ the time-gated SPAD receiver in OWC systems to mitigated the dead-time induced ISI effects. The statistics of the detected photon count of the time-gated SPAD receiver  are investigated. Due to the trade-off between the probability of detecting photon arrivals and the performance degradation induced by ISI, the gate-ON time interval can be optimized to achieve the best communication performance. Through extensive numerical results, it is demonstrated that compared to the traditional free-running receiver,  the time-gated receiver with optimal gate-ON time can significantly improve the BER performance and the background light tolerance especially in high received signal power regime.  

\section{Acknowledgements}
We gratefully acknowledge the financial support from EPSRC under grant EP/R023123/1 (ARROW).

\bibliographystyle{IEEEtran}
\bibliography{IEEEabrv,ECOC}

\end{document}